\documentclass{aa}  

\usepackage{graphicx}
\usepackage{txfonts}
\usepackage{orcidlink}

\defcitealias{Pallero26b}{P26}

\usepackage{xcolor}
\usepackage{hyperref}
\hypersetup{
    colorlinks=true,
    citecolor=blue,
    linkcolor=blue,
    filecolor=blue,      
    urlcolor=blue,
    }

\begin{document}

   \title{Characterizing the Formation and Evolution of S0-galaxies (CaFES-0): Revealing the origin of the mass-size relation for S0 galaxies.}
 
 \author{Diego Pallero\orcidlink{0000-0002-1577-7475}
          \inst{1,2}\fnmsep\thanks{\email{diego.pallero@usm.cl}}\and
          Yara~L. Jaff\'e\orcidlink{0000-0003-2150-1130}\inst{1,2}\and
          Amelia Fraser-McKelvie\orcidlink{0000-0001-9557-5648}\inst{3}\and
          Lodovico Coccato\orcidlink{0000-0001-7817-6995}\inst{3}\and
          Facundo A. Gómez\orcidlink{0000-0003-4232-8584}\inst{4}\and
          Evelyn J. Johnston\orcidlink{0000-0002-2368-6469}\inst{5}\and
          Ciria Lima-Dias\orcidlink{0009-0006-0373-8168}\inst{4}\and
          Maria Emilia De Rossi\orcidlink{0000-0002-4575-6886}\inst{6,7} \and
          Gissel P. Montaguth\orcidlink{0009-0003-1364-3590}\inst{8}}

    \institute{Departamento de Física, Universidad Técnica Federico Santa María, Avenida España 1680, Valparaíso, Chile \and
     Millennium Nucleus for Galaxies (MINGAL)\and
     European Southern Observatory, Karl-Schwarzschild-Straße 2,
    Garching, 85748, Germany \and
    Departamento de Astronom\'ia, Universidad de La Serena, Avda. Ra\'ul Bitr\'an 1305, La Serena, Chile \and Instituto de Estudios Astrof\'isicos, Facultad de Ingenier\'ia y Ciencias, Universidad Diego Portales, Av. Ej\'ercito Libertador 441, Santiago, Chile \and
    Universidad de Buenos Aires, Facultad de Ciencias Exactas y Naturales y Ciclo Básico Común, Buenos Aires, Argentina\and
    CONICET-Universidad de Buenos Aires, Instituto de Astronomía y Física del Espacio (IAFE), Buenos Aires, Argentina\and
    Departamento de Astronomia, Instituto de Astronomia, Geofísica e Ciências Atmosféricas da USP, Cidade Universitária, 05508-090 São Paulo, SP, Brazil}

   \date{Received September 15, 1996; accepted March 16, 1997}

  \abstract
{We investigate the structural evolution and formation pathways of lenticular (S0) galaxies using the Hydrangea suite of cosmological hydrodynamical simulations. Simulated galaxies reproduce the observed mass–size relation from the SAMI and MaNGA surveys, enabling a direct comparison between morphology, angular momentum, and size growth. We show that the S0 population occupies a characteristic V-shaped locus in the mass–size plane, which arises from the superposition of two physically distinct channels. Low-mass S0s are predominantly faded-formed S0s, quenched after infall into their present-day host halo and retaining the disk sizes of their star-forming progenitors. In contrast, high-mass S0s formed through mergers exhibit structural properties and size evolution similar to ellipticals, and typically quench before infall, consistent with pre-processing in group environments. By tracing their histories back to z = 1, we find that faded-formed S0s experience minimal structural evolution after quenching, whereas merger-formed S0s grow significantly in size through dissipationless interactions. These divergent evolutionary pathways explain both the slope break and the overall scatter of the S0 mass–size relation, demonstrating that lenticular galaxies arise from multiple formation mechanisms that leave distinct structural imprints.}

   \titlerunning{The mass-size relation of S0s}
\authorrunning{D. Pallero et al.}
   \maketitle
%

\section{Introduction}
\label{sec:intro}
One of the most enduring approaches to characterising galaxies is through their morphology, a classification scheme that reflects both their visual appearance and underlying physical structure. The traditional Hubble sequence \citep{Hubble26} distinguishes systems into ellipticals (E), spirals (S), and lenticulars (S0), with S0s occupying an intermediate position between featureless ellipticals and spirals with arms or bars. 
This framework was refined through more detailed subdivisions \citep{deVaucouleurs59}, including the recognition of barred variants (SB and SB0) as a distinct morphological axis within the sequence. Bars may influence the structural properties of lenticulars through secular processes such as gas funnelling and bulge growth \citep[e.g.,][]{Kormendy04, Athanassoula13, Frosst26}.
Recently, the classification has been further extended through kinematic \citep[e.g.,][]{Emsellem07, Emsellem11, Capellari11, Cappellari16} and photometric decompositions \citep[e.g.,][]{Conselice03, Simard11, Montaguth23, Lima-Dias21, Lima-Dias24, Sampaio25}, thereby providing a more physically grounded classification.

While morphology captures galaxy structure qualitatively, scaling relations provide quantitative, physically motivated constraints on how galaxies assemble their mass and evolve. Key examples include the mass–size relation \citep[e.g.,][]{Shen03, VanderWel09, VanderWel14, Lange15, Liao26}, the Tully–Fisher relation for rotationally supported galaxies \citep[e.g.,][]{Tully77, Bell01, Bamford06, Bedregal06}, and the Fundamental Plane for dispersion-supported systems \citep[e.g.,][]{Kormendy77, Dressler87, Capellari11, Romeo20, Romeo23}. Together, these empirical relations form a tight web of correlations linking stellar mass, size, luminosity, and kinematics. Lenticular galaxies consistently occupy intermediate or transitional loci within them \citep{VanderWel09, VanderWel14, Cortese16}, suggesting that S0s may reflect multiple evolutionary pathways rather than a single channel.

Among these relations, the evolution of galaxy sizes is particularly informative because it connects internal processes with the assembly of dark matter haloes.
Galaxies are generally expected to increase in size as they assemble mass, reflecting the cumulative imprint of their angular-momentum acquisition and evolutionary processes. In star-forming disk galaxies, new stellar mass forms at larger radii as fresh gas with higher specific angular momentum accretes, naturally driving an inside-out growth of the stellar disk. For quiescent and early-type systems, size growth is instead dominated by dissipationless processes, particularly minor mergers, which add stars to the outskirts and expand the effective radius more efficiently than they increase the total stellar mass. Together, these mechanisms produce a positive correlation between stellar mass and size and underlie the observed morphological differences and the redshift evolution of the mass–size relation.
Nonetheless, statistically speaking, the size–mass relation varies significantly with morphology \citep{Shen03, VanderWel14}: late-type galaxies follow a relatively shallow relation in which stellar density grows with mass, while early types display a more complex trend, showing a pronounced density peak near $M_\star \sim 4\times10^{10}$~M$_\odot$ and declines toward both higher and lower masses \citep{Kormendy77, Cappellari16}. Notably, deviations from these relations often signal transformative events such as mergers, compaction phases, or quenching processes. Thus, scaling relations collectively act as a fossil record, preserving the signatures of a galaxy’s growth history.

This connection between structure and evolution naturally leads to the broader question of the roles of internal (“nature”) and external (“nurture”) mechanisms in shaping galaxies. Initial conditions such as angular momentum and halo mass influence whether systems form as discs or spheroids \citep[e.g.,][]{Fall80, Mo98}, whereas environmental processes—including galaxy interactions, mergers, and cluster-specific mechanisms—can quench star formation and drive morphological transformation \citep[e.g.,][]{Dressler80, Dressler84, Boselli06, Cortese21}. Understanding how these competing influences act and interact across cosmic time is essential for interpreting modern scaling relations and the diversity of galaxy morphologies.

In this context, lenticular galaxies remain among the most challenging classes to interpret. They are broadly defined as disc galaxies that lack spiral arms and significant star formation, yet many display substantial structural subcomponents, such as bulges, bars, and rings. This broad and somewhat ambiguous definition reflects their diversity and hints at multiple evolutionary origins. Proposed formation channels include the fading of spiral galaxies through gas depletion \citep[e.g.,][]{Larson80, Bekki02}, minor mergers and tidal heating \citep[e.g.,][]{Bekki11, Tapia17, Deeley21}, and environmental quenching processes such as ram-pressure stripping \citep[e.g.,][]{deLucia12, Peng15, Pallero22}. However, the relative contributions of these mechanisms—and their imprints on scaling relations—remain open questions.

A significant step forward has come from state-of-the-art cosmological simulations, which now reproduce galaxy populations consistent with observed colours, masses, and sizes \citep[e.g.,][]{Trayford16, Furlong17, Nelson18, Genel18}. Large-volume hydrodynamical simulations such as EAGLE \citep{Schaye15, Trayford15}, IllustrisTNG \citep{Nelson18, Pillepich18, Naiman18, Marinacci18, Springel18}, and Horizon-AGN \citep{Dubois16} allow the tracking of morphological transformations over cosmic time and provide a controlled setting to disentangle the interplay between environment, stellar feedback, and mergers. Recent studies have begun to quantify the role of these processes in producing S0s and shaping their structural evolution \citep[e.g.,][]{Lagos17, Lagos18, Rositob, Rosito19a, Pallero19, Pallero22, Pallero26b}. In parallel, machine-learning approaches are enhancing the ability to identify morphological analogues across simulations and observations \citep[e.g.,][]{Huertas-Company19}.

In a companion study \citet[][P26 hereafter]{Pallero26b}, we investigated the formation pathways of S0 galaxies using the same Hydrangea simulation suite, focusing on how their merger histories and environmental histories determine their present-day morphology. In \citetalias{Pallero26b}, we show that the vast majority ($>85\%$) of S0s surrounding dense regions reside as satellites in massive haloes ($\log_{10} M_{200}/\mathrm{M}_\odot > 13$), 
and that $\sim75\%$ of these satellite S0s have undergone no significant mergers since $z=2$. Based on this, we classify S0s into two broad populations: \textit{faded-formed S0s}, which are satellite galaxies quenched by environmental mechanisms after infall (predominantly ram-pressure stripping and starvation), and \textit{merger-formed S0s}, which are typically central galaxies in 
lower-density environments that formed through at least one significant interaction with a mass ratio $f_i > 1:10$. Importantly, \citetalias{Pallero26b} demonstrates that S0 mergers preferentially occurred in low-density environments at $1 < z < 2$, prior to the infall of these galaxies into their current host haloes, supporting a pre-processing scenario in which morphological transformation is largely decoupled from the present-day cluster environment. These results, further validated by comparison with the observational classification of \citet{Coccato22}, provide the physical basis for the dual formation framework 
adopted in this study.

That said, in this work, we investigate how distinct formation pathways shape the positions of S0 galaxies in the mass–size plane using the Hydrangea cosmological simulations. By combining morpho-kinematic classification with star-formation activity, we identify S0s and trace their evolutionary histories to disentangle the roles of mergers and environmental quenching. We demonstrate that the characteristic V-shaped mass–size relation of S0 galaxies emerges from the superposition of two physically distinct channels: merger-driven growth and the fading of spiral progenitors. This study is part of the CaFES-0 project, which aims to systematically characterise the formation pathways of lenticular galaxies across environments.

This paper is structured as follows. Section~\ref{sec:data} presents the simulations and observational datasets, while Section~\ref {sec:methods} presents our methodology for measuring kinematic properties and the morpho-kinematic selection of our galaxies. Section~\ref{sec:results} analyzes the regions occupied by different morphological populations in the mass–size plane and evaluates the role of pre-processing. Section~\ref{sec:disc} discusses the implications for galaxy evolution across environments. Finally, Section~\ref{sec:summary} summarises our conclusions.

The simulation used on this work assumes the same flat $\Lambda$CDM cosmology adopted in the original \textsc{EAGLE} project \citep{Schaye15, Crain15}, specifically those of \citet{Planck14}, being: $\Omega_\Lambda = 0.693$, $\Omega_m = 0.307$, $\Omega_b = 0.04825$, $\sigma_8 = 0.8288$, $Y = 0.248$, and $H_0 = 67.77$ km s$^{-1}$ Mpc$^{-1}$.

\section{Data}
\label{sec:data}
\subsection{The \textsc{Hydrangea} simulations}
\label{sec:sims}
In this section, we provide an overview of the main characteristics of the Hydrangea simulations, a subset of the \textsc{C-EAGLE} simulation suite. We refer the reader to \citet{Barnes17} and \citet{Bahe17} for comprehensive details on the simulation framework and subgrid physics.

The C-EAGLE project comprises 30 high-resolution cosmological zoom-in simulations targeting massive galaxy clusters, with halo masses in the range $14 \leq \log_{10}(M_{200}^{z=0}/\mathrm{M}_\odot) \leq 15.4$ \citep{Barnes17}. These clusters were selected from a (3.2 Gpc)$^3$ parent dark matter-only simulation and resimulated with baryonic physics using the AGNdT9 variant of the EAGLE galaxy formation model \citep{Schaye15, Schaller15}, within a flat $\Lambda$CDM cosmology consistent with Planck 2014 parameters \citep{Planck14}.

All simulations adopt particle masses resolution of $m_{\mathrm{DM}} = 9.7 \times 10^6,\mathrm{M}\odot$ and $m{\mathrm{gas}} = 1.8 \times 10^6,\mathrm{M}\odot$, with a Plummer-equivalent gravitational softening of $\epsilon = 0.7$ kpc (physical) at $z < 2.8$. Zoom-in regions extend to at least 5 $r_{200c}$, and up to 10 $r_{200c}$ in the Hydrangea subset \citep{Bahe17}, enabling the study of galaxy evolution well beyond cluster cores.

Simulations were run with an enhanced version of the Tree-PM SPH code GADGET-3, incorporating the ANARCHY hydrodynamics scheme \citep{Schaye15}, which includes improvements in time integration, artificial viscosity, and conduction. Subgrid physics includes metal-line cooling \citep{Wiersma09}, a UV/X-ray background \citep{Haardt01}, stochastic star formation following a pressure-dependent Kennicutt–Schmidt law \citep{Schaye08}, and thermal supernova feedback \citep{DallaVecchia08}. Stellar evolution and chemical enrichment follow the treatment of \citet{Wiersma09b}, which tracks 11 key elements.

Halos and substructures are identified in each of the 30 output snapshots (from $z = 14$ to 0) using the SUBFIND algorithm \citep{Springel01}, preceded by a Friends-of-Friends (FoF) step. Only FoF groups with at least 32 dark matter particles are considered. To trace galaxy evolution across snapshots, the SPIDERWEB algorithm \citep{Bahe19} builds merger trees by tracking particle IDs, offering reliable galaxy tracking in dense cluster environments.

For this study, we will define galaxies as those subhaloes with $M_\star > 10^{10}$, ensuring sufficient resolution ($\gtrsim 10^4$ stellar particles per snapshot), located within 10 $r_{200}$ of each cluster centre to avoid contamination from low-resolution boundaries.

\subsection{The SAMI and MaNGA galaxy surveys}
\label{sec:sami}
The Sydney-AAO Multi-object Integral field spectrograph \citep[SAMI; ][]{SAMI} galaxy survey \citep{Croom21} observed 3068 galaxies in both the GAMA field regions and eight galaxy clusters.
The Mapping Nearby Galaxies with APO (MaNGA) galaxy survey \citep{Bundy15} observed $\sim 10000$ galaxies in the Sloan Digital Sky Survey fields. 

Here, we employ the S0 catalogue of \citet{Coccato22}, where S0s were selected visually using pre-existing morphology catalogues for each survey \citep[][]{Cortese16, Vazquez-Mata22}. For details on their classification, please refer to \citet{Coccato22}, which discusses potential biases introduced by visual vs kinematic selection techniques and sSFR-derived selection techniques in Appendix A.

\begin{figure*}

\includegraphics[width=\textwidth]{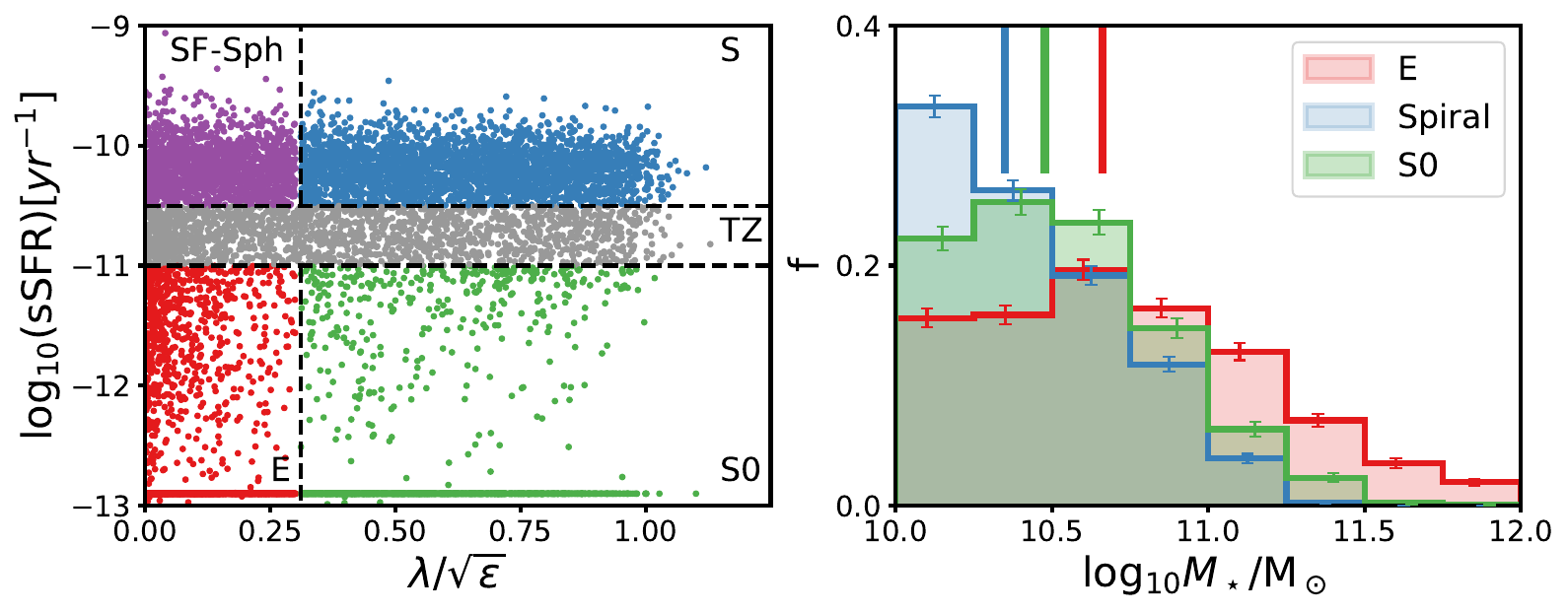}

\caption{ Left panel: Morphokinematic classification of galaxies based on their specific star formation rate and $\lambda/\sqrt{\varepsilon}$ ratio.
The dashed lines divide galaxies into five regions based on their kinematical support and sSFR. Galaxies are divided as spirals (star-forming disks), lenticulars (non-star-forming disks), star-forming ellipticals (star-forming spheroids), ellipticals (non-star-forming spheroids), and a transition zone ($-11 < \log_{10} \rm{sSFR}~[\rm{yr}^{-1}] < -10.5$). Right panel: Stellar mass distribution for the sample of E (red), S (blue), and S0 (green) galaxies. SF-Sph and transition galaxies are excluded from this panel. The lines at the top, colored the same, show the medians of each population. Error bars correspond to the binomial uncertainties associated with each bin.}

\label{fig:sel_all}
\end{figure*}

While both the SAMI and MaNGA surveys observed nearby ($z<0.15$) galaxies with fibre-fed integral field spectrographs of spaxel size 0.5'', these instruments were mounted on different telescopes and at different locations. Hence, the natural seeing and instrumental effects differ between the surveys. We attempt to homogenise derived global parameters by using consistent structural parameters and photometry to derive stellar masses. For galaxies observed by either the SAMI or MaNGA galaxy surveys, the effective radius, $R_e$, is drawn from the NASA Sloan Atlas \citep[NSA; ][]{Blanton11}. 
Stellar masses were taken from the GALEX-Sloan-WISE Legacy Catalogue-2 \citep[GSWLC-2; ][]{Salim16, Salim18} using a sky match with maximum allowed separation of 2 arcsec. In that catalogue, stellar masses were determined via SED fitting using the Code Investigating GALaxy Emission \citep[CIGALE; ][]{Noll09, Boquien19}. The photometry included in the fits was sourced from the Wide-Field Infrared Survey Explorer (WISE) in the mid-infrared, the Sloan Digital Sky Survey (SDSS) in the optical, and the Galaxy Evolution Explorer (GALEX) in the UV.

The specific angular momentum parameter within one effective radius, $\lambda_{\rm{Re}}$, was calculated following the method of \citep{Fraser-McKelvie18}. Briefly, it is defined by \citet{Emsellem07, Emsellem11} as
\begin{equation}
\lambda_{\rm{Re}} = \frac{\langle R|V| \rangle}{\langle R\sqrt{V^{2} + \sigma^{2}}\rangle} = \frac{\sum^{N}_{i=0} F_{i}R_{i}|V_{i}|}{\sum^{N}_{i=0}F_{i}R_{i}\sqrt{V_{i}^{2} + \sigma_{i}^{{2}}}}
\end{equation}

where $F$ is the flux, $V$ the stellar rotational velocity, and $\sigma$ the stellar velocity dispersion of the $i^{th}$ spaxel. In the same manner as Fraser-McKelvie et al. (2021), we define $R$ as the semimajor axis of an ellipse on which spaxel $i$ lies. $N$ is the total number of spaxels within 1 $Re$. Measurements are then corrected for inclination following \citet{Emsellem11}.

$\lambda_{\rm{Re}}$ is influenced by both the full width at half-maximum of the point spread function (PSF) (the `seeing' of the observation) and the inclination of the galaxy. We therefore correct them using the PSF correction of \citet{Harborne20}. Note that $\lambda_{\rm{Re}}$ is a flux-weighed quantity.

\section{Methods}
\label{sec:methods}
\subsection{Morphokinematic properties}
\label{sec:kin}

To classify galaxies in our simulation, we will employ a combination of the morpho-kinematic properties first defined in \citep{Emsellem07, Emsellem11} and then thoroughly tested on simulations \citep[e.g., ][]{Naab14, Lagos17, Pallero25}. 

In what follows, we summarise the main aspects of the methodology for defining the specific angular momentum parameter $\lambda$ in simulations, as described in \citet{Lagos17}.

To ensure consistency and minimize noise, all kinematic and structural properties are measured within a fixed aperture of three stellar half-mass radii ($r_{hsmr})$. This choice enables us to focus on the central, well-resolved regions of galaxies, allowing for robust comparisons across the sample.

We begin by computing the \textit{specific angular momentum} of the stellar component:

\begin{equation}
    j_{\star} = \frac{\sum_i m_i (\mathbf{r}_i - \mathbf{r}_{\rm COM}) \times (\mathbf{v}_i - \mathbf{v}_{\rm COM})}{\sum_i m_i}
\end{equation}

where $\mathbf{r}_i$ and $\mathbf{v}_i$ are the position and velocity of the $i$-th stellar particle, and $\mathbf{r}_{\rm COM}$, $\mathbf{v}_{\rm COM}$ denote the galaxy’s center of mass. From this, the rotational velocity is calculated as:

\begin{equation}
    V_{\rm rot}(r) = \frac{|j_\star(r)|}{r}
\end{equation}

To assess random motions, we compute the velocity dispersion perpendicular to the disk plane. Stellar velocities are projected along the angular momentum vector $\mathbf{L}_\star$, giving:

\begin{equation}
    \sigma_{1,D} = \sqrt{ \frac{ \sum_i m_i \left[ (\mathbf{v}_i - \mathbf{v}_{\rm COM}) \cdot \hat{L}_\star \right]^2 }{ \sum_i m_i } }
\end{equation}

where $\hat{L}_\star$ is the unit vector in the direction of $\mathbf{L}_\star$.

Finally, we quantify the \textit{$\lambda_R$} from the stellar mass as:

\begin{equation}
    \lambda_R = \frac{\sum_{i=1}^{N(r)} m_{\star,i} r_i V_{\rm rot}(r_i)}{\sum_{i=1}^{N(r)} m_{\star,i} \sqrt{V_{\rm rot}^{2}(r_i) + \sigma_{1D,\star}^{2}(r_i)}}
\end{equation}

calculated in concentric radial bins within the aperture, following \citet{Naab14}.

To measure a galaxy's \textit{ellipticity}, we first compute the inertia tensor components as:

\begin{equation}
    I_{xx} = \sum_i m_i (y_i^2 + z_i^2), \quad I_{xy} = -\sum_i m_i x_i y_i
\end{equation}

The full tensor is diagonalised to obtain eigenvalues ($\lambda_1 \leq \lambda_2 \leq \lambda_3$), from which the semi-axis lengths are derived as:

\begin{equation}
\begin{split}
    a &= \sqrt{\lambda_2 + \lambda_3 - \lambda_1} \\
    c &= \sqrt{\lambda_1 + \lambda_2 - \lambda_3}
\end{split}
\end{equation}

Finally, the ellipticity is defined as:

\begin{equation}
    \varepsilon = 1 - \frac{c}{a}
\end{equation}

where $a$ and $c$ correspond to the major and minor axes, respectively.

\subsection{Morphological classification}
\label{sec:morph}
\begin{figure*}
\centering
\includegraphics[width=\textwidth]{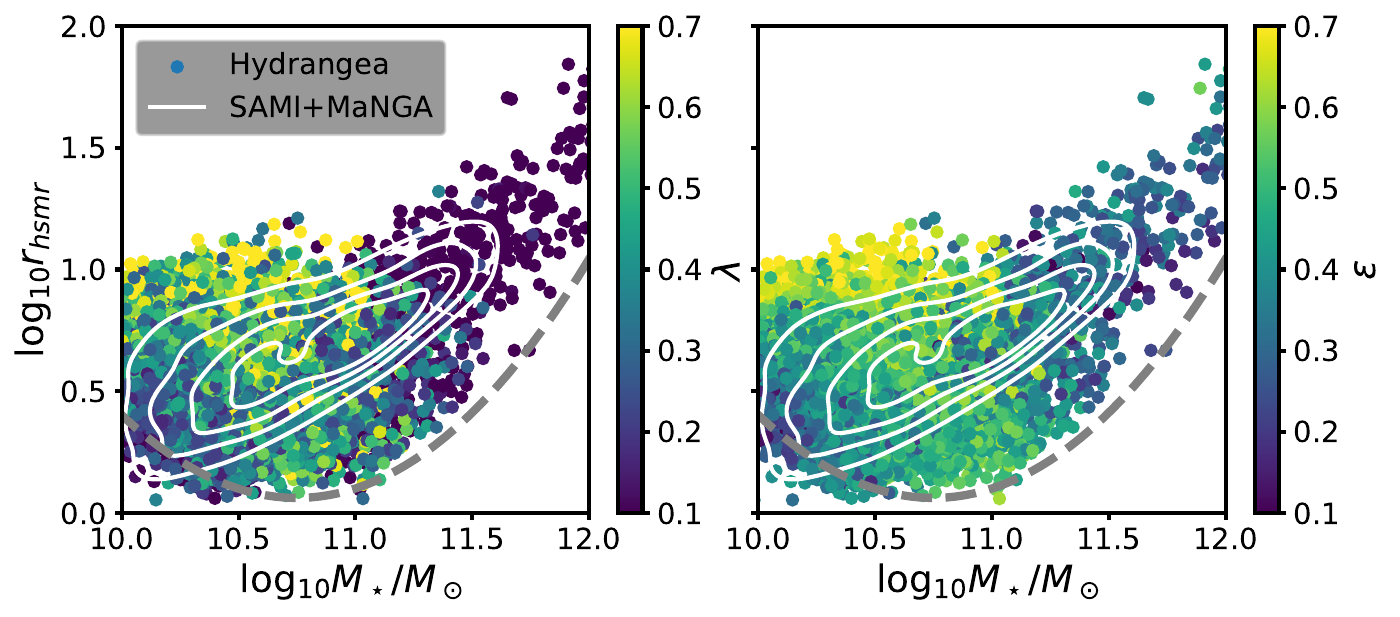}

\caption{Mass-size relation for all galaxies in the simulation, colour coded by their angular momentum, $\lambda$ (left panel), and their ellipticity, $\varepsilon$ (right panel). White contours show the mass-size relation for a sample of SAMI and MaNGA galaxies with different morphological types. Additionally, a grey dashed line indicates the "Zone of Avoidance`` (ZoA), a lower limit that includes 99\% of all early-type galaxies as defined in \citep{Cappellari13b}. 
As expected, galaxies grow in size with increasing mass, exhibit lower angular momentum, and have lower ellipticities. In the mass range of $10 < $log$_{10}M_\star/$M$_\odot < 11$, there is a mix between angular momentum and ellipticities, showing that the bulk of different morphologies is hidden in this region. Additionally, it is shown that although the sizes in observations correspond to the half-light radius, and in the simulation to the stellar half-mass radius, both samples follow the same quantitative relation. }

\label{fig:mass_size_all}
\end{figure*}

We will repeat the morphological classification performed in \citetalias{Pallero26b} for our simulated galaxies, combining morpho-kinematic properties and sSFR to split galaxies into S, E, and S0 types. 

That is, first we split ``spheroidal'' and ``disk'' galaxies using a threshold in $\lambda/\sqrt{\varepsilon}$ as proposed by \citet[][]{Emsellem07, Emsellem11}. In this form, we define galaxies as rotationally supported if $\lambda/\sqrt{\varepsilon} > 0.31$. On the other hand, galaxies with $\lambda/\sqrt{\varepsilon} < 0.31$ are defined as spheroids or slow rotators.

Additionally, we imposed a threshold on sSFR to select star-forming and quiescent systems, with log$_{10}$sSFR$ < -11 yr^{-1}$.
This threshold selection has been used in several other works to define galaxies without active star formation at $z \sim 0$ \citep[e.g.,][]{Muzzin12, Wetzel12, Dave19, Pallero19, Pallero22}. Although it is known that S0 galaxies are not necessarily fully quenched, we use this threshold to ensure that our S0 sample is as clean as possible, consistent with our classical definition. Nevertheless, it should be noted that, as discussed by \citet[][]{Pallero22}, the threshold for sSFR used to define passive galaxies has little impact on understanding their formation within a 0.5dex window.

Using these criteria, our sample naturally splits into four populations: star-forming disks (S), Quenched disks (S0s), quenched spheroids (E) and star-forming spheroids (SF-Sph). 
Additionally, we added a fifth population, called ``The transition zone,'' that includes galaxies that should be in the \textit{Green Valley} based on their sSFR ($-11 < $log$_{10}$sSFR $< -10.5$). It is worth noting that other authors have used this region to start defining galaxies as quenched \citep[see, e.g.][]{Brown17, Lacerna22, Palma25}. Nevertheless, as mentioned earlier, adjusting the limits of the transition zone did not alter our conclusions. 
Additionally, as discussed in \citep{Pallero25}, the selection threshold in $\lambda/\sqrt{\varepsilon}$ to split between disks and spheroids has little to no impact on our conclusions.

\begin{figure*}
\centering
\includegraphics[width=\textwidth]{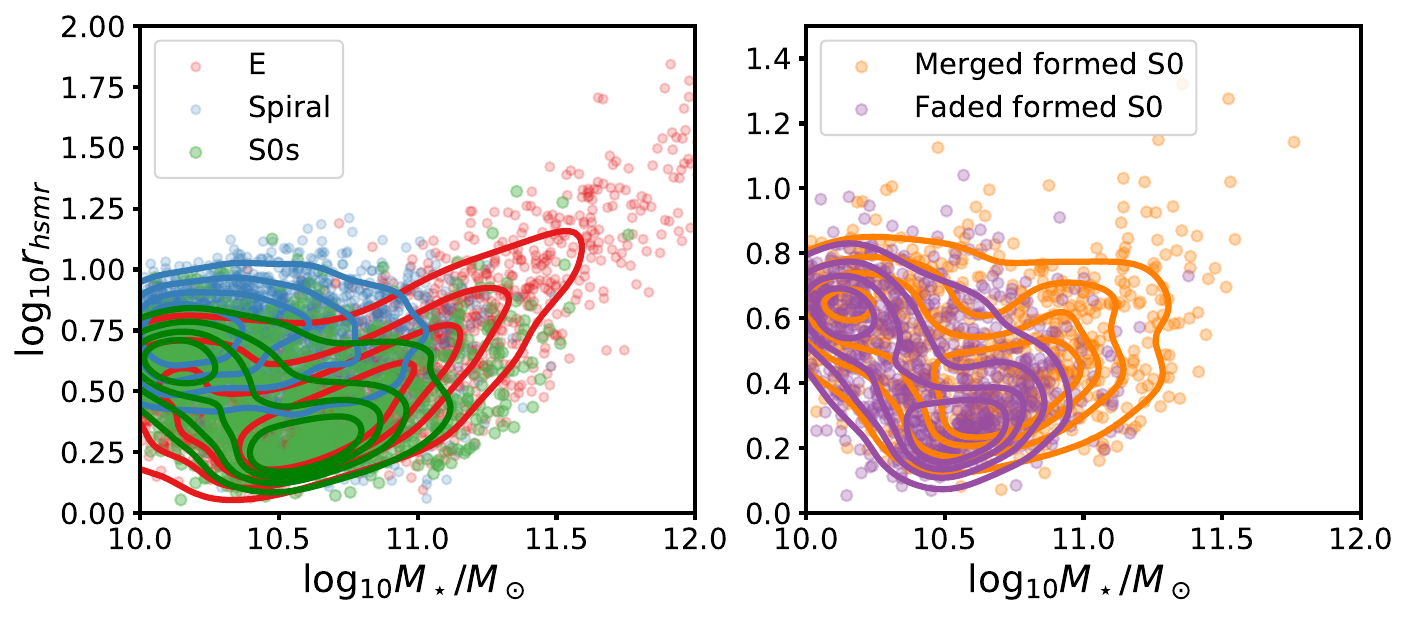}

\caption{Left: Mass-size relation for galaxies in the simulation split by morphological type.
Cyan, red, and magenta dots correspond to spiral, elliptical, and lenticular galaxies, respectively. Contours with the same colours are added to highlight the overall distribution of each galaxy population. Right: Mass-size relation for lenticular galaxies split by their formation mechanism. Orange and purple dots represent merged-formed and faded-formed lenticular galaxies, respectively.
Spiral galaxies show larger sizes for a given stellar mass, with a flat distribution in size, contrary to elliptical galaxies, which show smaller sizes at a given mass, but with a significant slope that increases their sizes significantly at log$_{10}M_\star/$M$_\odot > 11$. Lenticular galaxies, on the other hand, show an intermediate population with a negative slope at low masses with a sharp change in the slope of the distribution at log$_{10}M_\star/$M$_\odot = 10.5$, showing a V-like shape in their overall distribution. When splitting lenticular galaxies by their formation mechanisms, we see that the V-like shape is due to their different formation histories: faded-formed S0s show a negative slope (left part of the V-shape), while merged lenticular galaxies follow a distribution similar to that of elliptical galaxies (right part of the V-shape).
 }

\label{fig:mass_size_s0s}
\end{figure*}
The left panel of Figure \ref{fig:sel_all} shows the distribution of galaxies in our sample, split by their kinematical support and sSFR. As can be seen in this Figure, our entire sample of galaxies can be divided into:
\begin{itemize}
    \item Upper left quadrant (purple dots): Star-forming spheroids (log$_{10}$sSFR $> -10.5$ and $\lambda/\sqrt{\varepsilon} < 0.31$ ). Dwarf elliptical and irregular galaxies populate this region.
    \item Upper right quadrant (blue dots): Star-forming disks (log$_{10}$sSFR $> -10.5$ and $\lambda/\sqrt{\varepsilon} > 0.31$ ). Spiral galaxies (S) populate this region.
    \item Lower left quadrant (red dots): Quenched spheroids (log$_{10}$sSFR $< -11$ and $\lambda/\sqrt{\varepsilon} < 0.31$ ). Elliptical galaxies (E) populate this region.
    \item Lower right quadrant (green dots): Quenched disks (log$_{10}$sSFR $< -11$ and $\lambda/\sqrt{\varepsilon} > 0.31$ ). Lenticular (S0) galaxies populate this region.
    \item Central region (grey dots): Transition zone ($-11<$log$_{10}$sSFR $< -10.5$ ). Galaxies with a broad range of rotational support, transitioning from star-forming to quenched, populate this region.

\end{itemize}

From now on, we will work with samples of S, S0, and E galaxies to understand their distinct formation pathways. 

Additionally, the right panel of Figure \ref{fig:sel_all} shows the stellar mass distribution for E (red), S (blue), and S0 (green) galaxies. The median of each distribution is shown at the top of the panel, highlighting that each sample corresponds to a different stellar mass distribution, with S, S0, and E being more massive in order. To address this mass inequality between populations, in \citetalias{Pallero26b}, we create a control sample by selecting the same number of objects at each morphological type within the same stellar mass distribution. Nonetheless, for this work, we decided to keep the full sample of galaxies, as their positions in the mass-size plane reveal information that may be hidden by reducing the sample.

\section{Results}
\label{sec:results}
\subsection{The shape of the mass size relation}
\label{sec:msr}

Firstly, in Figure~\ref{fig:mass_size_all}, we aim to identify how our galaxy sample compares with a well-characterised observational sample from the SAMI and MaNGA Surveys. Galaxies in the simulation are colour-coded by their angular momentum (left panel) and ellipticity (right panel). White contours show the distribution of galaxies for the SAMI and MaNGA Surveys. Additionally, we include a grey dashed line showing the ``Zone of Avoidance'', corresponding to the lower envelope described by early-type galaxies including $99\%$ of the sample, as described in \citep{Cappellari13b}.
The simulated galaxies reproduce the observed mass--size relation despite the differences in how size is measured -- i.e., half-light radius in observations versus stellar half-mass radius in simulations -- and show a strong dependence on angular momentum and ellipticity.
As expected, galaxy size increases with stellar mass, with a clear dependence on internal kinematics. High-mass galaxies tend to have lower angular momentum and lower ellipticities, with 88\% and 98\% of galaxies with $\rm{log_{10}}M_\star/M_\odot > 11$ and 11.5, respectively, having $\lambda < 0.5$, consistent with their spheroidal morphologies. In the intermediate mass range (10 $\leq \rm{log_{10}}M_\star/$M$_\odot < 11$), we observe a broad distribution of angular momentum and ellipticity values, spanning a clear diversity in properties. Additionally, as shown in \citet{Cappellari13b}, at $M_\star \geq 3\times10^{11}M_\odot$ slow rotators dominate the mass-size plane, with the vast majority of galaxies having values of $\lambda \lesssim 0.1$.

\begin{figure}
\centering
\includegraphics[width=0.5\textwidth]{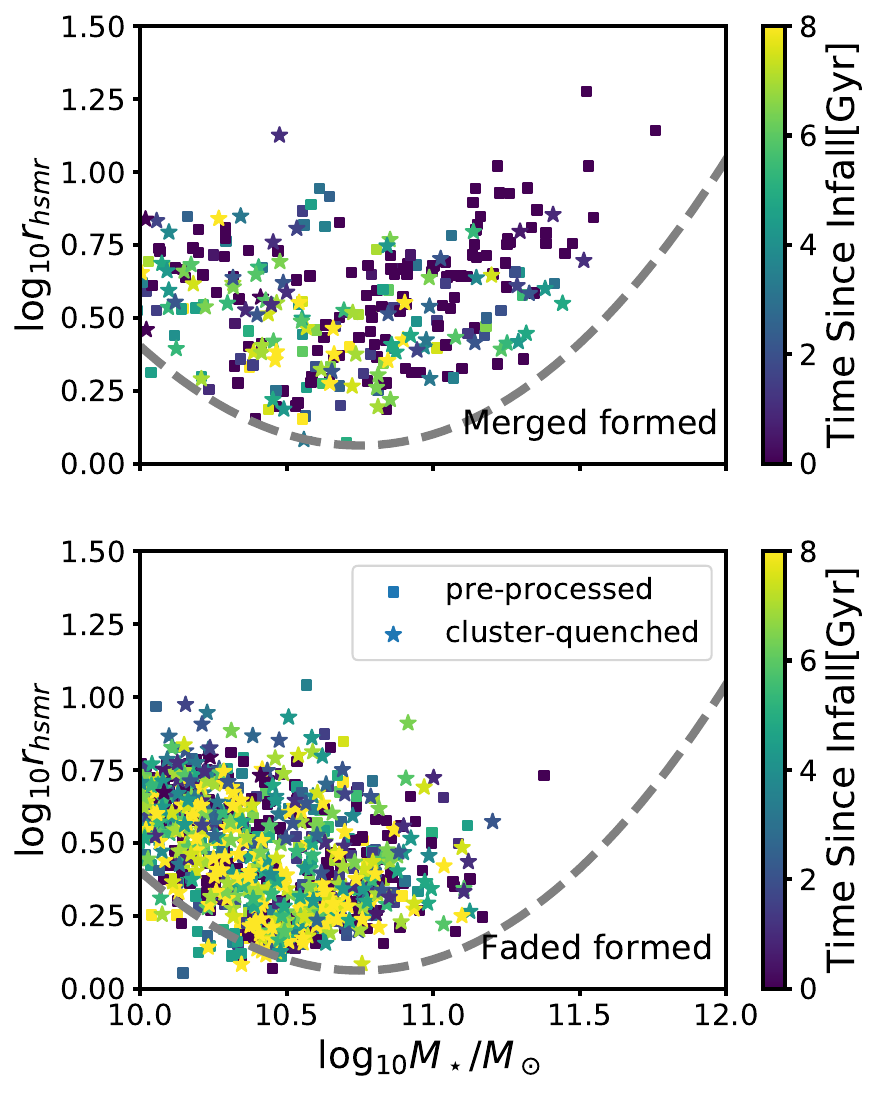}

\caption{Mass size relation for merged formed lenticular galaxies (upper panel) and faded-formed lenticular galaxies (lower panel), colour-coded by their time since infall to their host halo. Squares and stars correspond to pre-quenched and cluster-quenched galaxies, respectively, i.e., those that were quenched before their infall or after being accreted into their current host halo. The dashed grey line is the lower envelope, including 99\% of early-type galaxies, as a reference.
Faded-formed S0s are predominantly cluster-quenched galaxies; in contrast, pre-processed galaxies, which arrive already quenched, are primarily the result of merger-driven formation. Recently formed faded-formed S0s are larger, comparable in size to those of star-forming spirals. The most massive S0s are primarily merger-formed systems that have only recently fallen into the cluster.}

\label{fig:mstar_tsi}
\end{figure}

\begin{figure}
\centering
\includegraphics[width=0.5\textwidth]{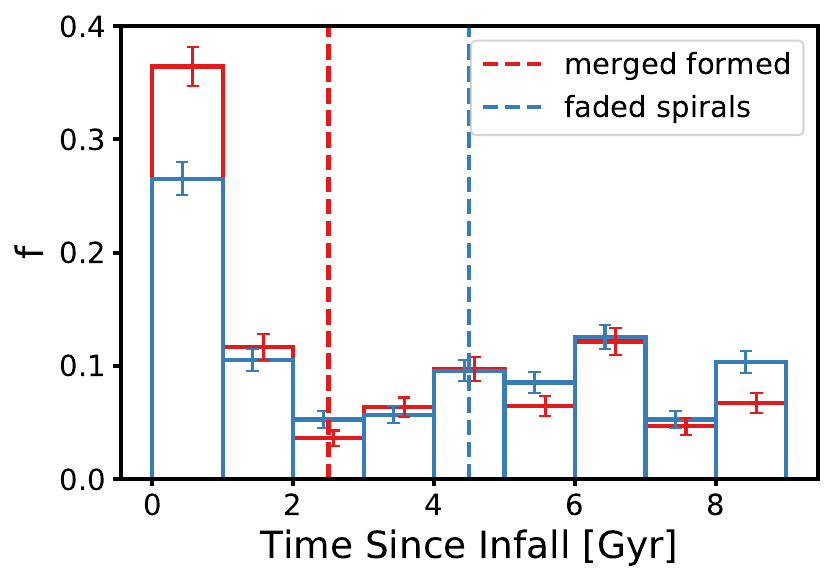}

\caption{Distribution of the time-since infall for satellite merged-formed lenticular galaxies (red) and faded-formed S0s (cyan). The dashed line shows the median distribution for each population. 
Error bars correspond to the binomial uncertainties associated with each bin.
Merged formed lenticulars have fallen onto their present-day host halo more recently than faded-formed S0s. When comparing the median, 
merged formed lenticular galaxies have a median distribution in Time Since Infall of $\sim 2.5$ Gyr ($ z=0.2$) while faded-formed S0s have a median of $\sim 4.5$ Gyr ($ z=0.5$).}

\label{fig:hist_tsi}
\end{figure}
These trends with mass become clearer when examining Figure \ref{fig:mass_size_s0s}, which further explores their relationship by separating the Hydrangea galaxy population by their morphokinematic classification. The left panel displays the mass–size relation for spiral (blue), elliptical (red), and lenticular (green) galaxies. Contours showing the $20^{th}, 40^{th}, 60^{th}$, and $80^{th}$ percentiles of the distribution for each population, for better visualisation.
Here, we can see that spirals exhibit the largest sizes at a fixed mass, with a relatively flat distribution as a function of stellar mass, similar to what is found from observational works \citep[e.g.,][]{VanderWel14}. Ellipticals are more compact, with a smoother V-shape trend than S0s, especially at intermediate masses, but show a steep positive slope at higher masses ($ M_\star/$M$_\odot \geq 11$), where size increases significantly. Lenticular galaxies lie intermediate in size and mass between the two, but their distribution forms a distinct V-shaped pattern with a slope transition at ($\approx M_\star/$M$_\odot \sim 10.5$).
Additionally, we can split our sample of lenticular galaxies based on their formation pathways. As mentioned above, two main formation pathways depend on the number of mergers S0s undergo. With this in mind, we split our S0s according to the number of significant mergers at $z\leq2$. We define a significant merger as any interaction in which the stellar mass ratio ($f_i$) between the interacting galaxies exceeds 1:10. As such, we split our sample between:
\begin{itemize}
    \item Faded formed: Galaxies that did not experience any significant merger at $z<2$
    \item Merged formed: Galaxies that experience, at least, one significant merger at $z<2$
\end{itemize}
The position that each population occupies in the size-mass relation is highlighted in the right panel of Figure~\ref{fig:mass_size_s0s}, where we additionally split the lenticular population by formation pathway: those formed through mergers (red) and those formed via fading of spiral galaxies (blue). This decomposition reveals that the V-shaped distribution arises from the superposition of two structurally distinct populations. Faded-formed S0s occupy the low-mass end of the lenticular population and exhibit larger sizes for less massive galaxies, with a negative slope.

\subsection{The role of pre-processing in shaping the mass-size relation}
\label{sec:prepro}
The accretion process of galaxies into their host haloes is inherently intertwined with their evolutionary history. In this sense, the role of pre-processing, i.e., the various physical mechanisms that can affect the evolution of galaxies before their accretion, is a crucial factor to consider~\citep{Fujita04}.
Throughout this article, we define ``pre-processed" as galaxies that arrive already quenched to their current halo. On the other hand, we will describe galaxies that reach their quenching state after infalling into their final host as ``cluster-quenched''.

\begin{figure*}
\centering
\includegraphics[width=\textwidth]{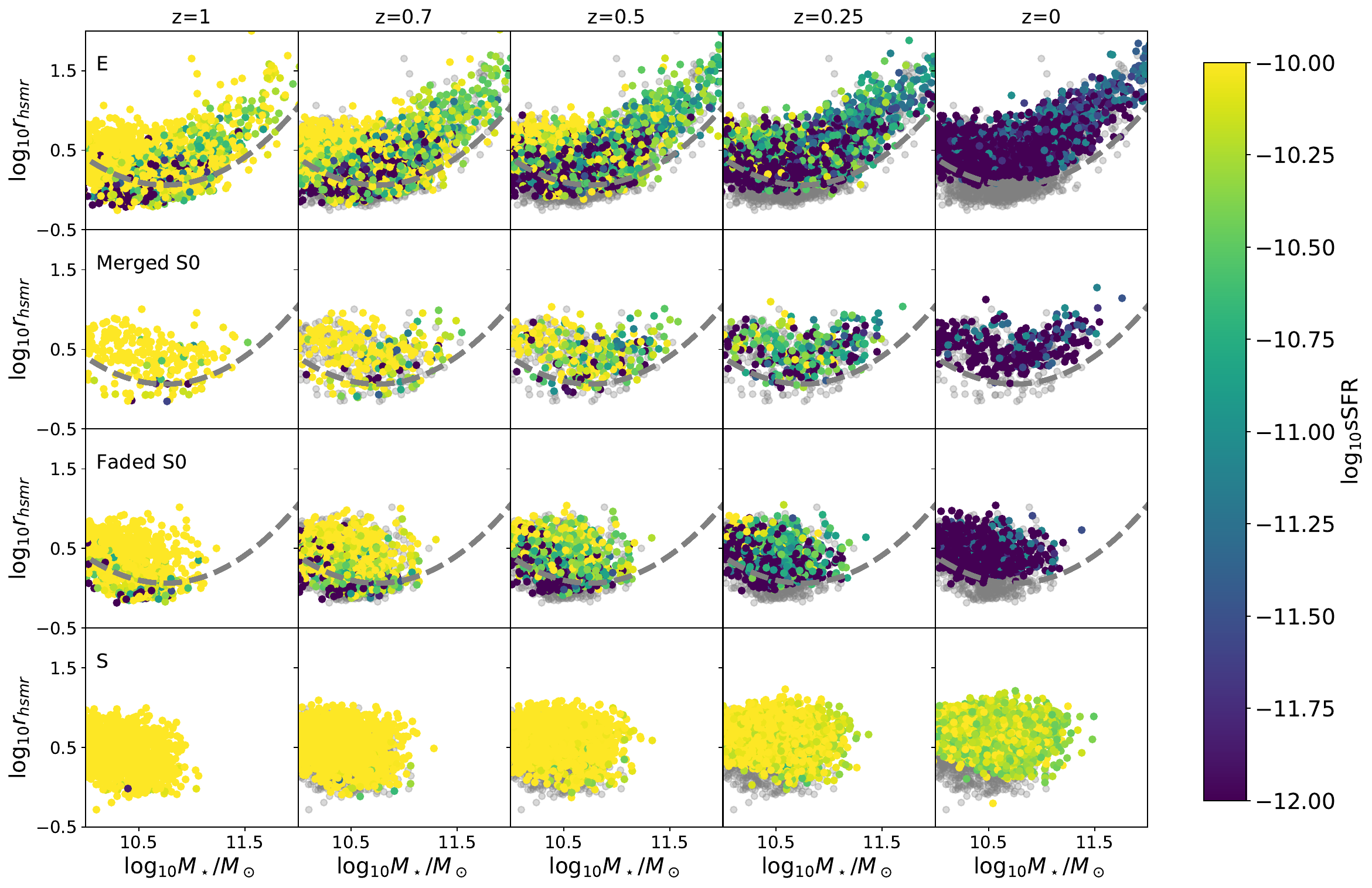}

\caption{Evolution in time for the mass size relation for galaxies selected at $z=0$ with different morphologies. From right to left, the mass-size relation is shown at z = 1, 0.7, 0.5, 0.25, and 0. The mass size relation for elliptical, merged-formed lenticulars, spirals, and faded-formed S0s is shown from the top to bottom panels. The lower envelope, including 99\% of all early-type galaxies at $z=0$, is shown in a grey dashed line as a reference. Galaxies are colour-coded by their sSFR at a given $z$. Grey dots in the background represent the mass--size relation at $z=1$ for visualisation purposes.} 

\label{fig:mass-size_evol}
\end{figure*}

\begin{figure*}
\centering
\includegraphics[width=0.9\textwidth]{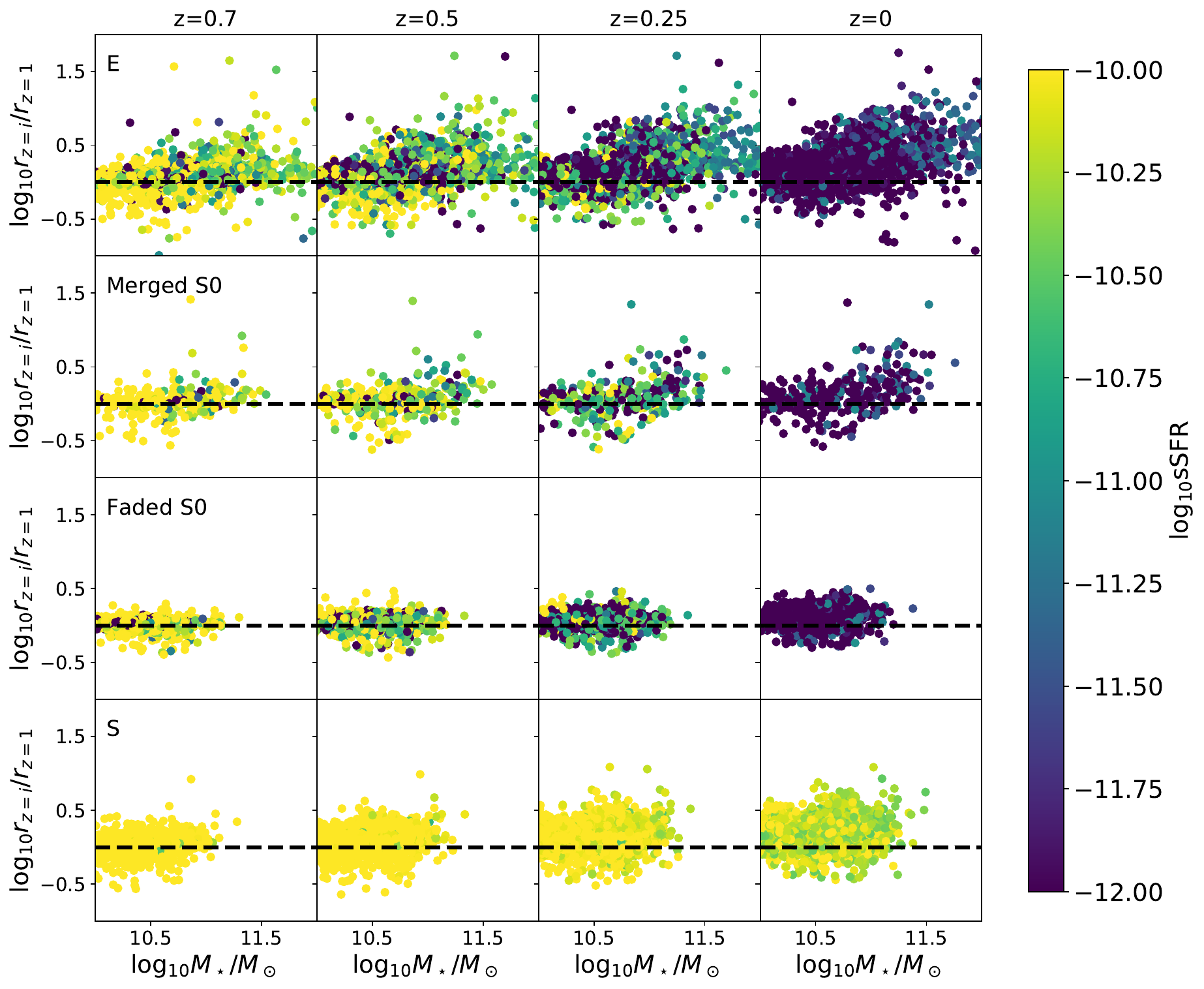}

\caption{Variations in size with respect to $z=1$ for galaxies with different morphologies as a function of their stellar mass. The results at z = 0.7, 0.5, 0.25, and 0, respectively, are shown from right to left. From top to bottom, we show elliptical, merged-form lenticulars, faded lenticulars, and spiral galaxies. Galaxies are colour-coded by their sSFR at a given redshift ($z$). The black dashed line on each panel shows log$r_z/r_{z=1} = 0$, i.e., when galaxies do not experience changes since $z=1$.}
\label{fig:mass-ratio_evol}
\end{figure*}

To understand how much time our galaxies have spent in their current environment, Figure~\ref{fig:mstar_tsi} shows the mass–size relation for lenticular galaxies separated by formation channel: merger-formed (top panel) and faded-formed S0s (bottom panel). Galaxies are colour-coded by their time since infall (TSI) into the cluster, where we define the infall time as the moment when a galaxy first crosses the $r_{200}$ of its host. Different symbols indicate whether they are pre-processed (squares) or cluster-quenched (stars).

Figure~\ref{fig:mstar_tsi} shows a clear trend: faded-formed S0s are predominantly cluster-quenched galaxies, indicating that their star formation was quenched due to environmental effects after accretion into the host halo. These galaxies often retain large sizes relative to their mass, particularly those that have been quenched recently, resembling the sizes of star-forming spiral galaxies. In contrast, merger-formed lenticulars are more likely to be pre-processed, having quenched before their accretion, and are generally more compact for a given stellar mass. However, the most massive S0s (log$_{10}M_\star/$M$_\odot > 11$) are almost exclusively merger-formed and typically have short infall times, suggesting they were accreted recently and are not subject to long-term cluster environmental effects.

This distinction is reinforced in Figure~\ref{fig:hist_tsi}, which shows the time distribution since infall (TSI) into their final host for the two populations. Faded S0s (blue) exhibit a broad and older distribution, with a median TSI of approximately 4.5 Gyr (corresponding to $z \sim 0.5$), indicating that they have been members of their corresponding host for more extended periods and have undergone prolonged environmental quenching. Conversely, merger-formed lenticulars (red) exhibit a more recent infall history, with a median TSI of $\sim$2.5 Gyr ($z\sim0.2$). These results are consistent with quenching occurring outside the host halo at $z=0$, with their quenching occurring outside the current host halo, and with their entering their final host as passive systems.

\subsection{The origin of the shape of the mass size relation}
\label{vshape}
To understand how the mass-size relation forms its shape, we followed the evolution of our galaxies, selected at $z=0$, up to $z=1$. This time, we split them by their present-day morphology to track how different formation pathways affect their positions in the mass-size relation.

\begin{figure}
\centering
\includegraphics[width=0.5\textwidth]{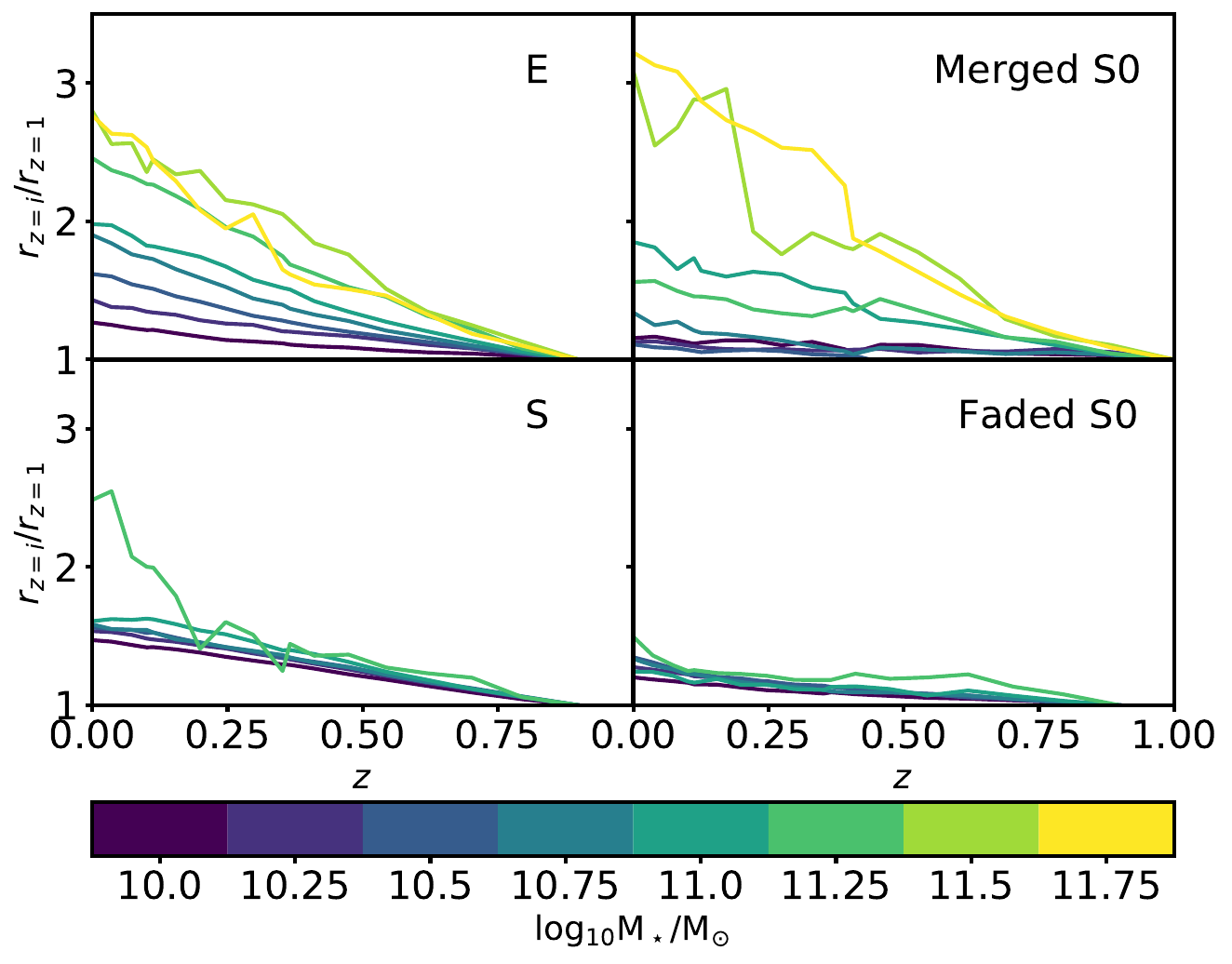}

\caption{Median growth in size of galaxies with different morphologies since $z=1$.
Galaxies are divided into eight stellar mass bins, selected at $z=0$, and colour-coded accordingly. Bins are 0.25dex wide between $10 \leq$ log$_{10}M_\star/$M$_\odot \leq 12$. Elliptical and merged-formed lenticular galaxies grow steadily since $z=1$, with more massive galaxies growing more than less massive galaxies.  
On the other hand, Spiral and faded spiral galaxies form lenticular galaxies, with no significant trend in stellar mass. Growing all the same, regardless of their stellar content. Moreover, faded-formed S0s show little to no growth since $z=1$ across all stellar mass bins.
}

\label{fig:median_mstar}
\end{figure}

\begin{figure}
\centering
\includegraphics[width=0.5\textwidth]{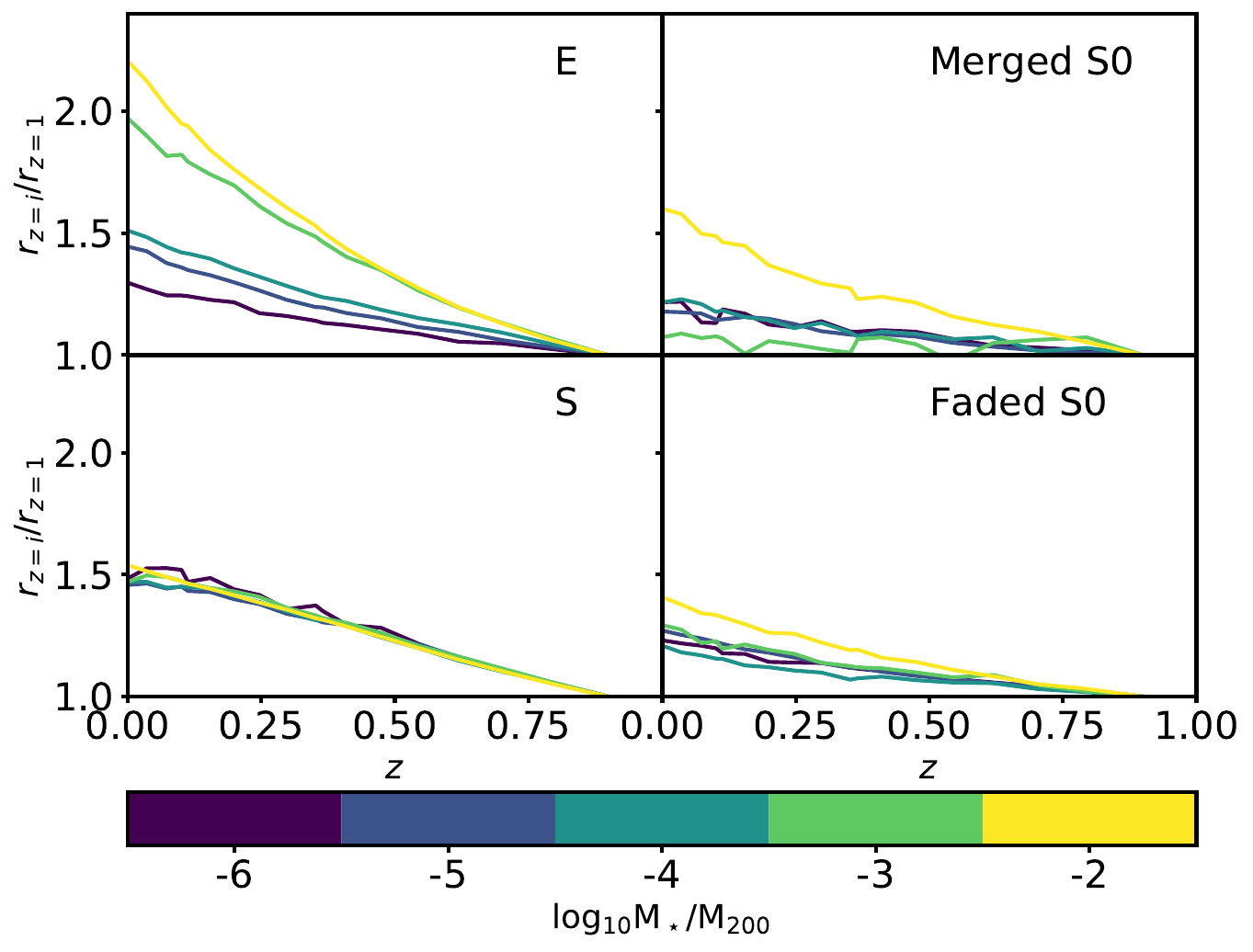}

\caption{Median growth in size of galaxies with different morphologies since $z=1$.
Galaxies are split into five bins of $M_\star/M_{200}$ selected at $z=0$ to trace how dominant the environment is for each galaxy.
Bins are one dex wide between $-6  \leq$ log$_{10}M_\star/$M$_{200} \leq -1$. 
Elliptical and merged-formed lenticular galaxies grow steadily, growing more when residing in haloes where they dominate.  
On the other hand, Spiral and faded-formed S0 galaxies show no significant trend with their environment. Growing all the same, regardless of their dominance in their respective environment. Moreover, faded-formed S0s show little to no growth since $z=1$.}

\label{fig:median_ratio}
\end{figure}
For this, Figure~\ref{fig:mass-size_evol} shows the mass–size relation at multiple redshifts ($ z=1,0.7,0.5,0.25,0$) from left to right, for ellipticals, merger-formed lenticulars, spirals, and faded-formed S0s (from top to bottom panels). Galaxies are colour-coded by their sSFR at each epoch. Grey dots represent the mass-size relation for each morphological type at $z=1$, highlighting temporal changes.

For spiral and elliptical galaxies, the relation evolves smoothly: both populations grow in size and stellar mass with time, while maintaining a consistent scatter. At a fixed mass, galaxies at lower redshifts tend to be larger, reflecting a gradual increase in size across cosmic time. This is particularly pronounced for ellipticals, which grow through mergers, especially at high redshift, as shown in \citetalias{Pallero26b}.

In contrast, lenticular galaxies exhibit more complex evolutionary patterns. Merged-formed S0s display size and mass growth similar to ellipticals, consistent with their merger-driven origin. However, faded-formed S0s experience only modest changes in size and mass since $z=1$, reflecting a more passive evolution following quenching. 

To quantify these changes, Figure~\ref{fig:mass-ratio_evol} shows the variations in galaxy size relative to $z=1$, i.e, $r_z/r_{z=1}$, as a function of stellar mass for the same galaxy populations at multiple redshifts ($ z=0.7,0.5,0.25,0$, from left to right). Panels are again arranged by morphology from top to bottom as in Figure \ref{fig:mass-size_evol}. Galaxies are colour-coded by their sSFR. The dashed line marks no size change since $z=1$ (i.e., $r_z/r_{z=1} = 1$).

A striking trend is seen for ellipticals and merger-formed S0s: at high redshift, they show a broad distribution of size changes, driven by frequent mergers. By lower redshifts, this distribution narrows, revealing a clear mass-dependent trend:
\begin{itemize}
    \item Low-mass galaxies (M$_\star \leq 10^{10.5}$M$_\odot$) tend to shrink relative to their sizes at $z=1$
    \item Massive galaxies show significant growth, particularly above $log_{10}M_\star/$M$_\odot \geq 11$
\end{itemize}

This dual behaviour strengthens the observed V-shaped mass–size relation at 
$z=0$, where the origin lies in the divergent size evolution of massive and low-mass early-type galaxies.

Meanwhile, spirals exhibit a consistent size growth across all stellar masses, in line with inside-out disk growth. Faded-formed S0s remain essentially unchanged since $z=1$, with minimal structural evolution ($<30\%$ in median compared to the $\sim300\%$ for massive E and merged-formed S0s). 

\section{Discussion}
\label{sec:disc}

In this section, we interpret the physical origin of the V-shaped mass–size relation of S0 galaxies, as shown in Section \ref{sec:results}, and discuss its implications for their formation pathways. Our results support a dual evolutionary scenario in which low-mass S0s arise from faded spiral galaxies quenched by environmental processes, whereas high-mass S0s primarily form through mergers that drive structural growth. In the following, we discuss the role of environment, merger history, and pre-processing in shaping these populations.

\subsection{Physical origin of the V-shaped mass--size relation}
\label{sec:vshape_origin}

The V-shaped locus occupied by S0 galaxies in the mass--size plane arises from the superposition of two structurally distinct populations with different formation histories, as shown in Section~\ref{sec:msr}. 
Below, we discuss the physical mechanisms responsible for the characteristic structural properties of each population.

Faded-formed S0s dominate the low-mass end of the S0 distribution and exhibit a negative slope in the mass--size plane — that is, less massive galaxies tend to be larger for their mass. This reflects their disk-dominated origins: 
These systems are quenched primarily by environmental processes, in particular ram-pressure stripping \citep[as shown in \citetalias{Pallero26b}; see also][]{Pallero22}, which depletes the gas reservoir from the outskirts inward, inhibiting further 
star formation and size growth at large radii while concentrating the stellar mass in the inner regions. As a result, faded-formed S0s show little to no structural transformation following quenching and retain the disk sizes of their spiral progenitors. This is further supported by their minimal evolution in size since $z=1$ (Section~\ref{vshape}), reinforcing the notion that these galaxies are passively evolved disks, structurally preserved despite their environmental transformation. We note, however, that fly-by encounters and harassment within galaxy clusters may also contribute to the evolution 
of some faded-formed S0s, potentially introducing scatter in their mass--size distribution.

In contrast, merger-formed S0s dominate the high-mass end and follow a structural trend closely resembling that of elliptical galaxies, with sizes increasing steeply above $\log_{10}M_\star/\mathrm{M}_\odot \gtrsim 10.5$. 
Their more compact sizes at intermediate masses, relative to faded-formed S0s, reflect the dissipationless nature of the mergers that assembled them, which efficiently redistribute stellar mass and grow the effective radius through the addition of stars at larger radii. This is consistent with their pre-processed nature reported in \citetalias{Pallero26b}, whereby these galaxies were quenched early in low-density group environments and subsequently accreted into their present-day host haloes as already passive systems, with their structural properties shaped primarily by their pre-infall merger history rather than by cluster-specific mechanisms.

Together, these two populations give rise to the observed V-shape: the left arm is defined by the negative mass--size slope of faded-formed S0s, while the right arm traces the elliptical-like growth of merger-formed S0s. 
The transition between the two branches near $\log_{10}M_\star/\mathrm{M}_\odot \approx 10.5$ marks the stellar mass scale at which merger-driven formation begins to dominate over environmental fading, and reflects the broader picture in which the two formation channels operate in different mass and environmental regimes.

\subsection{Nature vs nurture: size growth dependence on stellar mass and environment}
\label{sec:naturevsnurture}

Figure~\ref{fig:median_mstar} presents the median size growth for ellipticals, merger-formed lenticulars, spirals, and faded-formed lenticulars across eight stellar mass bins from $log_{10}M_\star/$M$_\odot = 10$ to 12. Ellipticals and merger-formed S0s exhibit progressive, mass-dependent size growth since $z=1$, with more massive systems growing more significantly---consistent with more frequent and impactful mergers in massive early-type galaxies. For the low-mass population ($log_{10}M_\star/$M$_\odot$ < 10.5), galaxies can even shrink relative to $z=1$, a behaviour absent at higher masses. In contrast, spirals and faded-fr+lr show minimal and roughly uniform growth across all mass bins, with faded-formed S0s displaying little to no size evolution since $z=1$ regardless of stellar mass.
Figure~\ref{fig:median_ratio} complements this by analysing size growth as a function of halo dominance ($log_{10}M_\star/M_{200}$). Ellipticals and merger-formed S0s again show stronger growth at higher $log_{10}M_\star/M_{200}$, reinforcing the picture from \citealt{Pallero25} in which centrally dominant galaxies experience more mergers and grow in both size and mass. This result is also consistent with the scenario proposed by \citet{Romeo20, Romeo23}, in which a galaxy's status as a central or satellite can either foster or hinder its formation and long-term survival. 
Spirals and faded-formed S0s show no clear dependence on halo position, indicating that secular processes---not environment---govern their structural evolution.
Together, these results reveal a fundamental dichotomy: merger-formed S0s and ellipticals undergo environment- and mass-dependent structural growth, while faded-formed S0s evolve passively, preserving their structural properties after quenching as fossil disks. Although both S0 populations may appear morphologically similar at $z=0$, their divergent evolutionary histories mark them as physically distinct classes of objects.

\subsection{Caveats}
\label{sec:caveats}

\subsubsection{Cluster-centric and stellar mass selection bias}

The Hydrangea suite targets massive clusters in the range $14 \leq \log_{10}(M_{200}^{z=0}/\mathrm{M}_\odot) \leq 15.4$, so our sample is inherently biased towards high-density environments. Although the zoom-in regions extend to $10\,r_{200}$, our conclusions about the relative importance of merger-driven and fading channels may not apply to S0s in lower-density environments such as isolated groups or the field \citep[e.g.,][]{Just10, Wilman09, Wilman12}. Future work extending this analysis to field and group environments will be important to test the universality of the V-shaped mass--size relation and the dual formation scenario described here.
Additionally, it is worth noting that our sample is limited to galaxies with $M_\star > 10^{10}$~M$_\odot$ to ensure 
sufficient numerical resolution ($\gtrsim 10^4$ stellar particles). This cut excludes the low-mass end of the lenticular population, where additional formation channels, such as harassment and tidal stripping, may play a more significant role \citep[e.g.,][]{Boselli06}. The negative slope in the mass--size relation for faded-formed S0s is therefore constrained to a relatively narrow mass range, and extending this trend to lower masses would require higher-resolution simulations and observations.

\subsubsection{Morpho-kinematic measurements and classification thresholds}

Our morphological classification relies on fixed thresholds in $\lambda/\sqrt{\varepsilon}$ and sSFR (Section~\ref{sec:morph}). While these have been calibrated in previous work \citep{Emsellem11, Lagos17, Pallero25} and shown to be robust within a 0.5~dex window in sSFR, the strict threshold ($\log_{10}~\mathrm{sSFR} < -11~\mathrm{yr}^{-1}$) may exclude green-valley galaxies that would otherwise be classified as lenticular on morphological grounds. Relaxing these thresholds does not qualitatively alter our conclusions \citep{Pallero22, Pallero25}.
Additionally, we compare stellar half-mass radii from the simulation with half-light radii from the SAMI and MaNGA surveys. The former is generally expected to be smaller than the latter, particularly in galaxies with strong colour gradients or young stellar populations at large radii \citep[e.g.,][]{Suess22}. While both samples follow a consistent mass--size relation (Section~\ref{sec:results}), this offset introduces a systematic uncertainty in the normalisation that would require synthetic photometry to fully resolve, which is beyond the scope of this work.

\subsubsection{Merger classification boundary}
We define merged S0s and faded-formed S0s based on whether a galaxy experienced at least one significant merger with a stellar mass ratio $f_i \geq 1:10$ at $z \leq 2$. This threshold is necessarily a simplification: interactions with mass ratios just below this limit may still induce morphological perturbations and structural changes, particularly in low-mass systems. 
Furthermore, merger trees are constructed using the \textsc{SPIDERWEB} algorithm, which tracks particle IDs. Unresolved or poorly tracked interactions in dense environments could therefore lead to the misclassification of a small fraction of galaxies.
We therefore caution that the boundary between these two populations is not sharp, and some contamination between samples is expected. Future analyses using higher-resolution simulations will help to better constrain this classification.

\section{Conclusions}
\label{sec:summary}

In this work, we analysed the mass–size relation and structural evolution of galaxies with different morphological types using a cosmological hydrodynamical simulation. We focused in particular on lenticular (S0) galaxies, which we classify into two populations based on their dominant formation channel: mergers (merged S0s) and fading of spiral galaxies (faded S0s). Our main findings are summarised below:

\begin{itemize}
    \item Simulated galaxies exhibit a well-defined mass–size relation, consistent with observational trends from SAMI and MaNGA. Angular momentum and ellipticity correlate with stellar mass and morphology. While spiral galaxies occupy the high-size, high-angular momentum regime, ellipticals populate the low-size, low-angular momentum region. S0s bridge these populations, displaying a characteristic "V-shaped" mass–size distribution driven by the superposition of the two formation pathways. Faded-formed S0s dominate the low-mass end with decreasing size for increasing in mass, while merged S0s resemble ellipticals at higher masses.

    \item We find a strong link between the formation mechanism of S0s and their infall histories. Faded-formed S0s are predominantly quenched after infall (cluster-quenched), while merged S0s are mainly quenched before infall (pre-processed). The time since infall distribution shows that merged S0s typically fell into their present host halos more recently (median $\sim2.5$ Gyr ago) than faded-formed S0s (median $\sim$4.5 Gyr ago). The most massive S0s tend to be remnants of recent mergers.

    \item Spiral and elliptical galaxies show significant size and mass growth since $z=1$, with more massive ellipticals experiencing more rapid growth. In contrast, S0s exhibit divergent behaviours. Merged S0s follow a growth pattern similar to ellipticals, while faded-formed S0s undergo minimal structural evolution. This dichotomy contributes to the distinct shape of the mass–size relation for S0 galaxies.

    \item We find that the size growth of ellipticals and merged S0s correlates with both stellar mass and environmental dominance. Massive, more dominant galaxies exhibit greater size evolution, likely driven by merger activity. In contrast, spiral galaxies and faded-formed S0s show no significant trend with either mass or environment. Additionally, our findings indicate that the structural evolution of faded-formed S0s is minimal over the past $\sim8$ Gyr.

\end{itemize}

Finally, our results highlight the distinct formation pathways of different lenticular galaxies and emphasise the importance of separating them by formation channel when analysing their structural properties. While merger-driven S0s resemble ellipticals in their evolutionary paths, faded-formed S0s retain the imprint of their disky progenitors. This dual origin, despite similar visual and kinematic morphologies, explains much of the observed diversity among S0s and reinforces the role of both environment and internal processes in shaping galaxy morphology.

\begin{acknowledgements}
      We want to thank the referee for their helpful and insightful comments, which significantly improved this work. We would also like to thank Rory Smith, Ulrike Kuchner and Yannick Bahé for their useful and insightful discussions. 
      DP and YLJ acknowledge support from Comité Mixto ESO-GOBIERNO DE CHILE.
      DP and YLJ acknowledge support from the Agencia Nacional de Investigaci\'on y Desarrollo (ANID) through Basal project FB210003, FONDECYT Regular projects 1241426 and 123044, and  Millennium Science Initiative Program NCN2024\_112. DP, AFM, and LC acknowledge support from the ESO Early-Career Scientific Visitor Programme for the development of this project.
      FAG acknowledges support from the ANID BASAL project FB210003, from the ANID FONDECYT Regular grant 1251493, and from the HORIZON-MSCA-2021-SE-01 Research and Innovation Programme under the Marie Sklodowska-Curie grant agreement number 101086388.
      E.J.~Johnston acknowledges the support from the ANID CATA-BASAL project FB210003.
      CL-D acknowledges a grant from the Agencia Nacional de Investigación y Desarrollo (ANID) through Fondecyt project 3250511.
      G.M. gratefully acknowledges the Fundação de Amparo à Pesquisa do Estado de São Paulo (FAPESP) for the support grant 2024/10923-3 and 2025/14602-0
\end{acknowledgements}

%
%

\bibliographystyle{aa} 
\bibliography{quench_final} 

\end{document}